\documentclass[aps,prstper,twocolumn,groupedaddress,showpacs,floatfix,10pt]{revtex4-1}
\usepackage{times,amssymb,amsmath}
\usepackage{array}
\usepackage{dcolumn}
\usepackage{ifpdf}
\usepackage{graphicx}
\usepackage{epstopdf}
\usepackage{tabularx}
\usepackage[%
  pdfstartview=={FitH},
  pdftitle={IPS of 2-4 ML NiO/Ag(001)},
  pdfauthor={Konrad Gillmeister},%
  bookmarks=true,
  hyperfigures=true,
  unicode=true,
  colorlinks=true,
  linkcolor=black,	
  citecolor=black,	
  urlcolor=black,	
]{hyperref}

\mathchardef\mhyphen="2D

\begin{document}


\title{Image potential states of ultrathin NiO films: A time-resolved two-photon photoemission study}
\author{K. Gillmeister}
\affiliation{Institut f\"{u}r Physik, Martin-Luther-Universit\"{a}t Halle-Wittenberg, Halle, Germany}
\author{M. Kiel}
\affiliation{Institut f\"{u}r Physik, Martin-Luther-Universit\"{a}t Halle-Wittenberg, Halle, Germany}
\author{W. Widdra}
\affiliation{Institut f\"{u}r Physik, Martin-Luther-Universit\"{a}t Halle-Wittenberg, Halle, Germany}
\affiliation{Max-Planck-Institut f{\"u}r Mikrostrukturphysik, Halle, Germany}
\date{\today}
\begin{abstract}
Well-ordered ultrathin films of NiO have been prepared on an Ag(001) substrate using molecular beam epitaxy. With the help of angle-resolved two-photon photoemission (2PPE) a series of image potential states (IPS) for film thicknesses of 2--4 monolayers (ML) has been identified. By time-resolved 2PPE, the lifetimes of the first three IPS and their dependence on the oxide film thickness have been determined. While the   lifetimes of the (n=1) IPS are all in the range of 27--42 fs, the values for the (n=2) IPS decrease from 85 fs for 2 ML to 33 fs for 4 ML. These differences are discussed in terms of a coupling to the layer-dependent electronic structure of the NiO ultrathin films.
\end{abstract}
\pacs{...}

\maketitle
\section{Introduction}
  Image potential states (IPS) form a special class of surface states, in which electrons are bound in front of a polarizable surface. As response of such near-surface electrons, polarization charges are induced at the surface-region of the bulk. They result in an attractive image potential that can bind the electrons perpendicularly to the surface. 
  
  Since the first substantiation of image potential states by 2PPE in 1985 \cite{Giesen1985}, it is nowadays the experimental method of choice for such investigations. IPS play a decisive role as model systems to study fundamental decay processes of excited electrons. Extensive research has been conducted on clean metal surfaces \cite{Schuppler1990,Schoenlein1991,Fauster1995,Shumay1998,Schaefer2000,Link2000,Marks2011}, rare gas adlayers on different metallic substrates \cite{Padowitz1992,Wolf1996,Reuss1999,Berthold2004,Guedde2005,Hotzel2007,Damm2009}, and organic thin films on metal surfaces \cite{Ge1999,Gaffney2001,Zhu2004,Kirchmann2005,Varene2011}. 
  
  Theoretically, the IPS are well understood and it is also possible to calculate the dynamics of electrons in IPS \cite{Echenique2004,Tsirkin2013,Marinica2002}. Image potential states of clean metal surfaces can be described using the so called bulk penetration model. Starting point for this model is that the decay processes of electrons in IPS are influenced by their interaction with bulk electrons \cite{Fauster1995}. Simply speaking, the larger the lifetime of an electron in an IPS is the lower is the penetration of its wavefunction into the bulk, and vice versa. In addition, one has to consider the influence of the long-ranging Coulomb interaction \cite{Echenique2000,Echenique2004}. It influences the available phase space for decay processes strongly.
  
  Image potential states of metal surfaces can be decoupled with the help of insulating adsorbates. Thereby, the decoupling takes place only for adlayers that do not introduce any electronic states in the energy region of the IPS. Monitoring the lifetime of the IPS of such systems yields values that can be one or two orders of magnitude higher than for clean metal surfaces depending on the morphology and electronic structure of the adlayers \cite{Berthold2004}. 
  
  Up to now, only a few 2PPE investigations on metal oxides have been reported whereas none of these investigations focus on the image potential states \cite{Onda2004,Tisdale2008,Takahashi2013,Deinert2014}. Here we present 2PPE results obtained for NiO(001) ultrathin films on Ag(001). For 2, 3, and 4 ML, long-living unoccupied features are observed near the vacuum level. They are characterized as image potential states.
  For well-ordered NiO films of different thicknesses, the dynamics of the IPS directly reflect the coupling to unoccupied NiO thin film states. With our investigations on NiO, the knowledge about image potential states can be extended to the region in between highly polarizable metal surfaces and insulating adlayers.

\section{Experimental Details\label{sec:exp}}
  Sample preparation and all experiments have been performed in an ultra-high vacuum (UHV) chamber working at a base pressure of $1\cdot10^{-10}$ mbar. The chamber is equipped with an optics for low energy electron diffraction (LEED), an X-ray source for X-ray photoelectron spectroscopy (XPS) as well as a UV source for ultraviolet photoelectron spectroscopy (UPS). The photoemitted electrons are detected by a hemispherical analyzer (Phoibos 150, SPECS, Germany) which contains a CCD detector for simultaneous recording of electron energy and emission angle. For 2PPE, a broadly tunable femtosecond laser system is used that includes two non-collinear optical parametric amplifiers (NOPAs) which are pumped by a 20 W all-fibre laser (IMPULSE, Clark) as described elsewhere \cite{Duncker2012}. For the experiments described here, UV laser pulses with an energy of 4.17--4.27 eV are combined with infrared (IR) laser pulses (1.68--1.72 eV) and have typical pulse lengths of 50--90 fs and 35--45 fs, respectively. 
  
    \begin{figure*}[htbp]
     \centering%
     \includegraphics[height=9.1cm, trim=0 0 0 0]{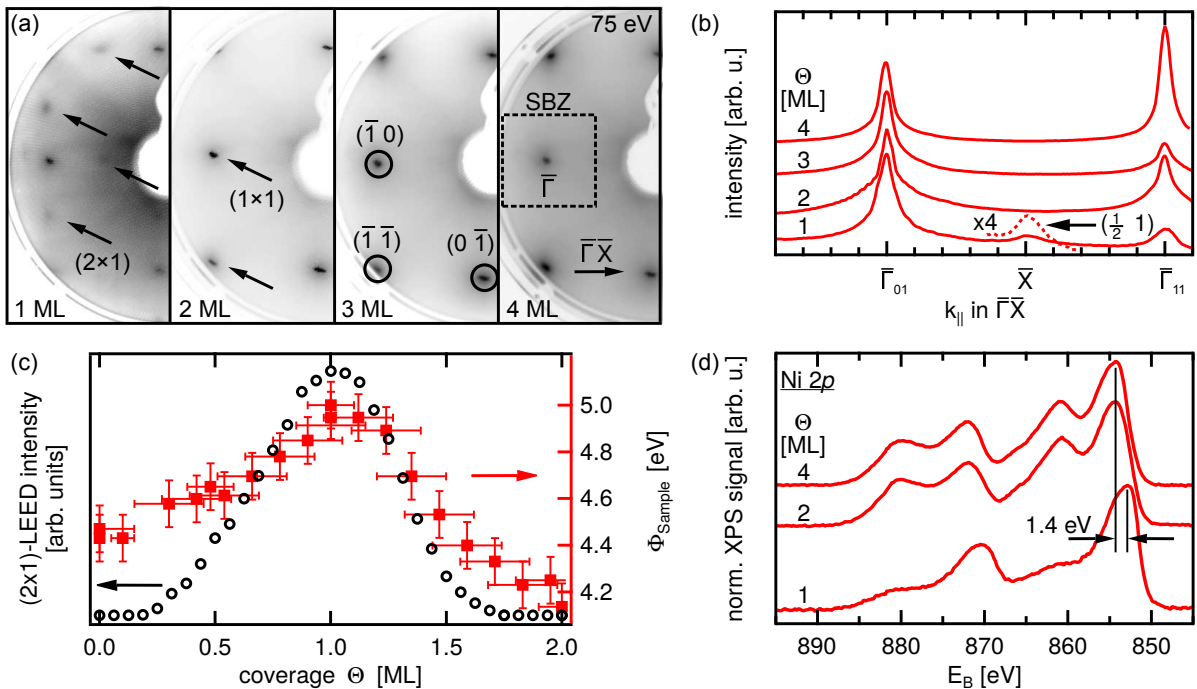}%
     \caption[]{(Color online) (a) Diffraction patterns of annealed NiOfilms on Ag(001) with thicknesses of 1--4 ML at 75 eV electron energy. The appearing diffraction spots, the surface Brillouin zone (SBZ) as well as the $\overline{\Gamma}\overline{X}$ direction are denoted. (b) Intensity profiles across the LEED images along the $\overline{\Gamma}\overline{X}$ direction of the SBZ. (c) LEED intensity of the $(2\times1)$ spots (black open circles, left scale) and workfunction $\Phi$ (red solid squares, right scale), respectively, vs. NiO coverage. (d) XP spectra of the Ni2$p$ core levels for 1, 2, and 4 ML NiO/Ag(001). All spectra are normalized to their maximum and shifted in intensity for the sake of clarity.}\label{pic:LEEDXPS}
    \end{figure*}%

  The Ag(001) crystal was cleaned by several cycles of Ar ion sputtering and subsequent annealing at 600 K until no contamination could be detected by XPS. NiO ultrathin films have been grown via molecular beam epitaxy (MBE) as established by Marre and Neddermeyer \cite{Marre1993}. Thereby, Ni was evaporated from a Ta crucible in an oxygen atmosphere of $2\cdot10^{-6}$ mbar with a total growth rate of 0.5 ML per minute. NiO films of 2, 3, and 4 ML thickness have been prepared and subsequently annealed at 540 K to improve long-range order and film perfection. The quality of the films has been monitored by the sharpness of the LEED spots (cf. Fig. \ref{pic:LEEDXPS} (a)). As reported also earlier, we find a layer-by-layer growth for NiO on Ag(001) \cite{Bertrams1996,Sebastian1999,Dhaka2011} for thicknesses from 2 ML and beyond. Hence, after calibrating the evaporator it was possible to prepare well-ordered NiO layers with an accuracy of $\pm10$ \% in thickness.
  
  A combination of LEED, XPS, and workfunction measurements has been used  for calibration of the NiO film thickness. In the monolayer regime, NiO shows a $(2\times1)$ structure \cite{Bertrams1996,Sebastian1999,Caffio2004,Caffio2006,Großer2008}. It is formed by a uniaxially distorted, nearly planar, quasihexagonal NiO structure \cite{Caffio2004,Grosser2008b}. Above 1 ML, the $(2\times1)$ structure vanishes and a pseudomorphic $(1\times1)$ film is formed \cite{Caffio2003}. In Fig. \ref{pic:LEEDXPS} (a), diffraction patterns of 1--4 ML NiO/Ag(001) are displayed. Intensity profiles along the $\overline{\Gamma}\overline{X}$ direction give clear evidence that the $(2\times1)$ diffraction spots only appear for the NiO monolayer (Fig. \ref{pic:LEEDXPS} (b)). Figure \ref{pic:LEEDXPS} (c) depicts the intensity of the $(2\times1)$ LEED spots with increasing coverage (left axis). For the closed monolayer, the $(2\times1)$ intensity is maximal. The structural transition from a quasihexagonal monolayer to rocksalt-like NiO films starting from 2 ML on leads to the disappearance of the $(2\times1)$ spots in LEED. 
    
  Due to the quasihexagonality, the chemical environment as well as the resulting surface dipole of the NiO monolayer is different from those of thicker films. These issues lead to unique features in photoelectron spectra \cite{Caffio2003,Caffio2004} and to a high workfunction with respect to other coverages. In Fig. \ref{pic:LEEDXPS} (d) XP spectra of the Ni 2$p$ core level are depicted for NiO films from 1--4 ML thickness. While the spectra for 2 and 4 ML are similar and even bulk-like in shape \cite{Alders1996,Sangaletti1997}, the spectrum for 1 ML NiO/Ag(001) differs significantly: The maximum of the Ni $2p_{3/2}$ peak appears with a difference of 1.4 eV at lower binding energy, and the satellite structure is differently shaped in comparison to the spectra for higher coverages. This is in accordance with findings in the literature \cite{Caffio2003,Caffio2004} and can be taken as a marker for this thickness.
  
  In Fig. \ref{pic:LEEDXPS} (c) the workfunction is plotted in dependence of the oxide film thickness. A further characteristic for 1 ML NiO/Ag(001) is the relatively high workfunction $\Phi_{\text{ML}}\sim5.0$ eV which differs strongly from that of clean Ag(001) ($\Phi_{\text{Ag}}\sim4.5$ eV) and these of thicker films ($\Phi_{\text{NiO}}\sim4.2...4.4$ eV). Therefore, one can clearly identify one monolayer thick films on the basis of workfunction measurements. 

\section{Results}
  
   \begin{figure}[htb]
     \centering%
     \includegraphics[width=0.96\linewidth, trim=0 0 0 0]{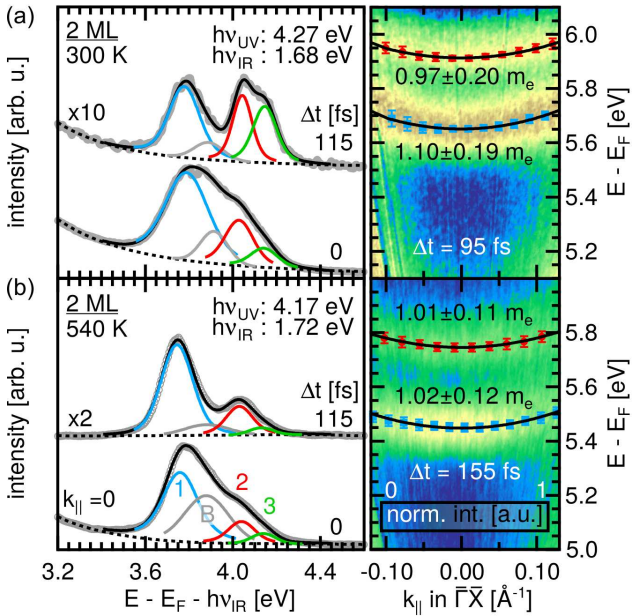}%
     \caption[]{(Color online) 2 ML NiO/Ag(001): 2PPE spectra at $k_{\parallel}=0$ (left panel) and momentum-resolved data (right panel). The data have been measured (a) directly after film preparation and (b) after annealing at 540 K for time delays of $\Delta$t=0 and $\Delta$t=115 fs, respectively. The as prepared (annealed) sample was pumped with h$\nu_{\text{UV}}$ = 4.27 eV (4.17 eV) and probed with h$\nu_{\text{IR}}$ = 1.68 eV (1.72 eV).}\label{pic:states2ML}
    \end{figure}
  
    \begin{figure}[htb]
     \centering%
     \includegraphics[width=0.96\linewidth, trim=0 0 0 0]{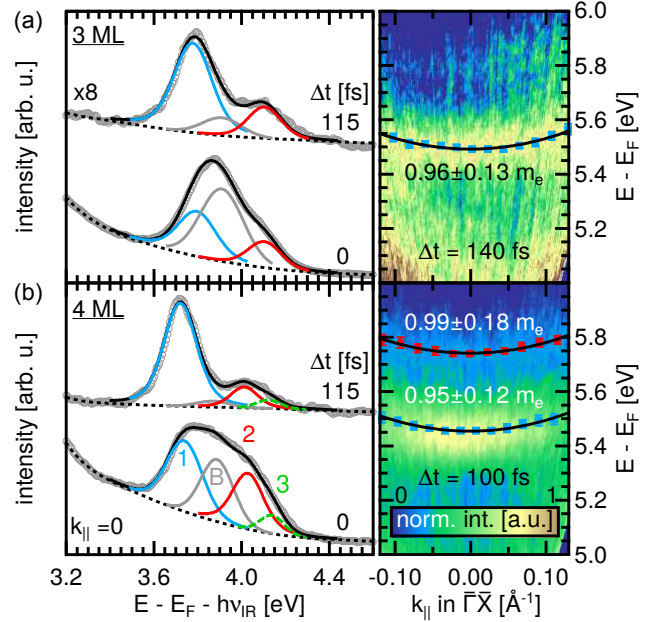}%
     \caption[]{(Color online) 2PPE spectra for $k_{\parallel}=0$ (left panel) and momentum-resolved data (right panel) for 3 ML (a) and 4 ML (b) NiO/Ag(001), pumped with h$\nu_{\text{UV}}$ = 4.17 eV and probed with h$\nu_{\text{IR}}$ = 1.72 eV at different time delays $\Delta$t.}\label{pic:states}
    \end{figure}

  In Fig. \ref{pic:states2ML} 2PPE spectra for two different preparations of 2 ML NiO/Ag(001) are depicted for fixed time delays $\Delta$t between pump (UV) and probe (IR) pulse. The spectra of Fig. \ref{pic:states2ML} (a) were obtained directly after film preparation at room temperature (RT) using photon energies of 4.27 eV and 1.68 eV for pump and probe, respectively. Four different features can be found in the spectra for time delays of $\Delta$t=0 and 115 fs. Long-living states dominate the spectrum at $\Delta$t=115 fs which are marked with 1, 2, and 3 in Fig. \ref{pic:states2ML} (b). These features correspond to intermediate state energies of 3.78, 4.04, and 4.14 eV (each $\pm$0.05 eV) above the Fermi level. A fourth short-living feature, tagged with B, is located at 3.91 eV. 
  
  The spectra of Fig. \ref{pic:states2ML} (b) were measured after annealing the sample at 540 K. Again, the spectra can be described by four different states at intermediate state energies of 3.75, 4.05, 4.14, and 3.90 eV (B). In contrast to the measurement directly after RT preparation, the spectra are dominated by feature 1 for both time delays $\Delta$t. Due to the reduced pump photon energy of 4.17 eV, state 3 is hardly populated and not visible in the spectra.
  
  Momentum-resolved data (right panel in Fig. \ref{pic:states2ML}) reveal a dispersion of the states 1 and 2 with effective masses for the as prepared (annealed) sample of $m_{e\!f\!f}\,=$ 1.1 (1.02) $m_e$ for feature 1 and 0.97 (1.01) $m_e$ for feature 2, respectively. This dispersion corresponds to that of a free electron in front of a surface. Therefore, the unoccupied states 1 and 2 are assigned to the (n=1) and (n=2) IPS of the oxide films. 
  
  For 3 and 4 ML NiO/Ag(001), similar unoccupied states are found (Fig. \ref{pic:states}). The intermediate state energies of the features are 3.79, 4.11 for the first and second IPS and 3.90 eV (B) for 3ML. They shift to 3.73, 4.02 (1, 2) and 3.89 eV (B) for 4 ML (each $\pm0.05$ eV). The energetic positions of these unoccupied states differ only slightly from those of the 2 ML states. According to their dispersion (right panel in Fig. \ref{pic:states}) with effective masses $m_{e\!f\!f}$ of 0.96, 0.95 (feature 1), and 0.99 $m_e$ (feature 2), the states are again assigned to (n=1) and (n=2) IPS, respectively. In addition, one can recognize a short-living feature B as has been seen for 2 ML. 
  
  In Table \ref{tab:positions}, these data are summarized together with the measured sample workfunctions $\Phi$. For the as prepared bilayer, the workfunction $\Phi$ amounts to 4.18 eV and is about 250 meV lower than for the annealed NiO films with 2--4 ML thickness. This difference is explained according to the Smoluchowski effect \cite{Smoluchowski1941} by a higher step density for the as prepared NiO film.

\begin{figure}[htb]
     \centering%
     \includegraphics[width=0.974\linewidth, trim=0 0 0 0]{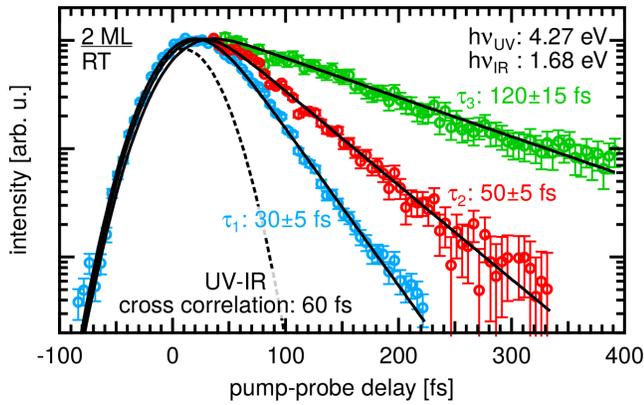}%
     \caption[]{(Color online) Time-dependent 2PPE intensities (data points) of the (n=1) IPS (blue, bottom), (n=2) IPS (red, middle), and (n=3) IPS (green, top) for the as prepared 2 ML NiO/Ag(001). The intensities are described by rate equations resulting in the decay curves (black solid lines) which yield the denoted well-defined lifetimes. The dashed black line represents the cross correlation (CC) between UV and IR pulses.}\label{pic:tau2ML}
    \end{figure}

  In Fig. \ref{pic:tau2ML}, time-dependent 2PPE intensities for the first three IPS are depicted for the as prepared 2 ML NiO(001) film. The data points are fitted using a rate equation approach (black solid lines). For comparison, the cross correlation curve between pump and probe pulse with a full width at half maximum (FWHM) of 60 fs is also included in Fig. \ref{pic:tau2ML}. The solid lines describe the experimental data well over two and a half order of magnitude in intensity and yield lifetimes of 30$\pm$5, 50$\pm$5, and 120$\pm$15 fs for the first three IPS. 
  
\begin{figure}[tb]
     \centering%
     \includegraphics[width=0.95\linewidth, trim=0 0 0 0]{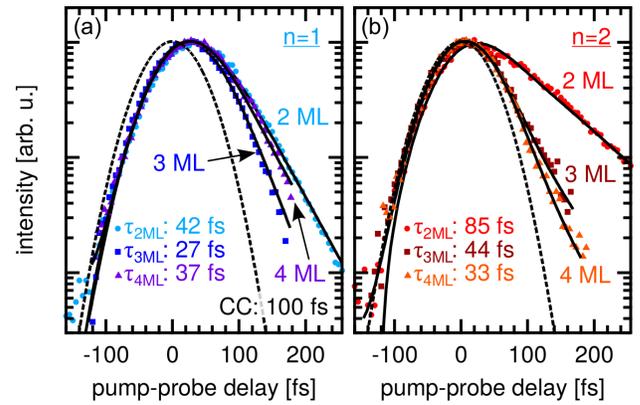}%
     \caption[]{(Color online) Time-dependent 2PPE intensities (data points) of a) the (n=1) IPS (blue colors) and b) the (n=2) IPS (red colors) of 2--4 ML NiO/Ag(001). For comparison, the CC traces are depicted as dashed black lines.}\label{pic:tau}
    \end{figure}

  For the \textit{annealed} 2 ML film, the lifetimes (Fig. \ref{pic:tau}) differ significantly from those of the as prepared film. The lifetime of the (n=1) IPS is about 10 fs longer (42$\pm$4 fs) whereas the lifetime of the (n=2) IPS has increased by more than 30 fs (85$\pm$15 fs). In Fig. \ref{pic:tau}, a comparison of the time-resolved 2PPE data for 2, 3, and 4 ML (all annealed to 540 K) is presented. The fits yield lifetimes of 27 and 37 fs (each $\pm$4 fs) for the (n=1) IPS of 3 and 4 ML. The lifetimes of the (n=2) IPS for 3 and 4 ML amount to 44$\pm$6 and 33$\pm$7 fs, respectively. 
  
\section{Discussion}
  
    In Table \ref{tab:positions}, the workfunctions $\Phi$, binding energies $E_B$, effective masses $m_{e\!f\!f}$, and lifetimes $\tau$ of the first three IPS are summarized for 2--4 ML NiO(001) on Ag(001). For the annealed NiO layers, the binding energies are in the range of 640--690 (320--390) meV for the first (second) IPS. The difference of 50 meV for the (n=1) IPS and 70 meV for the (n=2) IPS is close to the uncertainty of the measurement. Nevertheless, these variations can be explained by a slight modification of the NiO electronic structure with film thickness. 
    
    Compared to the annealed samples, the binding energies of the IPS of the as prepared sample as well as the workfunction are approximately 250 meV lower. In contrast, the energetic positions of the IPS \textit{with respect to the Fermi level} are almost unchanged for the as prepared as well as the annealed bilayer (cf. Fig. \ref{pic:states2ML}). This is interpreted within the concept of the local workfunction \cite{Fischer1993}. Workfunction measurements average over a large sample area containing flat NiO terraces as well as stepped NiO regions, resulting in a measured \textit{global} workfunction. Electrons in an IPS are located in front of the flat NiO terraces only. Therefore, they are affected by the \textit{local} workfunction of the flat NiO regions. Since the intermediate state energies of the IPS do not change for the as prepared and annealed NiO bilayer, the local workfunction does not change for both preparations. Due to an enhanced film quality upon annealing the NiO bilayer, the global workfunction is close to that of the flat terraces for the annealed NiO bilayer. Hence, the correct binding energy of the IPS is reflected by the annealed bilayer.
    
   \begin{table*}[tbp]
   \caption[]{Workfunctions $\Phi$, binding energies $E_B$, effective masses $m_{e\!f\!f}$, and lifetimes $\tau$ of the image potential states for 2--4 ML NiO/Ag(001).}
    \begin{ruledtabular}
     \begin{tabular}{ l c c c c c l c c c c }
     \multicolumn{1}{c}{\rule{0pt}{10pt}coverage}	& \multicolumn{1}{c}{$\Phi$} 	& \multicolumn{3}{c}{n=1}	& \multicolumn{3}{c}{n=2}	& \multicolumn{2}{c}{n=3}\\
      \cline{3-5} \cline{6-8} \cline{9-10} \rule{0pt}{12pt}
      \parbox[0pt][1.5em][c]{0cm}{}\text{[ML]}  & [eV] & $E_B$ [meV] & $m_{e\!f\!f}$ [$m_e$] & $\tau$ [fs] & $E_B$ [meV] & $m_{e\!f\!f}$ [$m_e$] & $\tau$ [fs] & $E_B$ [meV] & $\tau$ [fs]\\
     \hline	
     \rule{0pt}{12pt}
      2 (A)		& 4.18 & 400$\pm$100	& 1.10$\pm$0.19 & 30$\pm$5 & 120$\pm$100 & 0.95$\pm$0.20 & 50$\pm$5  & 40$\pm$40 	& 120$\pm$15\\
     \rule{0pt}{8pt}
      2 (B)		& 4.44 & 690$\pm$100	& 1.02$\pm$0.12 & 42$\pm$4 & 390$\pm$100 & 1.01$\pm$0.11 & 85$\pm$15 & 		& \\
     \rule{0pt}{8pt}
      3  		& 4.43 & 640$\pm$100	& 0.96$\pm$0.13 & 27$\pm$4 & 320$\pm$100 & 		   & 44$\pm$6 &			& \\
     \rule{0pt}{8pt}
      4 		& 4.41 & 680$\pm$100	& 0.95$\pm$0.12 & 37$\pm$4 & 390$\pm$100 & 0.99$\pm$0.18 & 33$\pm$7 &			& \\
    \end{tabular}\label{tab:positions}
    \end{ruledtabular}
  \end{table*}   
      
      The lifetimes of electrons excited into the (n=1) IPS (cf. Table \ref{tab:positions}) are for all coverages in the range of 27--42 fs. Here, the 12 fs longer lifetime for the 2 ML film annealed to 540 K (B) in comparison to the as prepared 2 ML (A) is explained by the higher film quality upon annealing. As known from other systems \cite{Schuppler1990,Reuss1999,Weinelt1999,Roth2002}, the electron dynamics strongly depend on defect and step densities. These densities are reduced upon annealing and lead to a reduced non-momentum conserving scattering. For the annealed oxide films, the quality of the layers of different thicknesses as determined by the sharpness of the LEED spots (cf. Fig. \ref{pic:LEEDXPS} (a)) is comparable. The film quality is significantly higher as compared to the as prepared 2 ML film. Therefore, the lifetimes in the annealed films should be less limited by surface imperfections.
    
    The relaxation of an IPS electron is dominantly dependent on the coupling of its wavefunction to wavefunctions of the substrate below. In our case, the latter corresponds to the electronic structure of the NiO(001)/Ag(001) thin film system. The remarkably shorter lifetime of electrons in the (n=1) IPS for 3 ML NiO/Ag(001) is, therefore, assigned to a stronger coupling to unoccupied NiO thin film states at this specific thickness. The different IPS lifetimes are, therefore, explainable by a slight modification of the electronic structure for 3 ML is slightly changed as compared to that of 2 and 4 ML thick NiO films. This is also substanciated by the different binding energies of the IPS for 3 ML as discussed above. 
    
    The decay of the (n=2) IPS behaves differently for the 2--4 ML films. First, one finds again a longer lifetime of the annealed 2 ML film as compared to 2 ML prepared at RT. This is explained by the improved film quality upon annealing. Clearly, the second IPS is even more sensitive to the film quality due to its longer lifetime as compared to the (n=1) IPS. The (n=2) IPS has an additional decay channel into the (n=1) IPS for non-momentum conserved scattering which makes it more defect sensitive. The decay times of the (n=2) IPS for the NiO bilayer are longer than those for the (n=1) IPS. This follows the general trend known from clean metals that the lifetime of an IPS is increased with increasing quantum number $n$ \cite{Echenique1978}. 
    
    The relaxation time of electrons from the (n=2) IPS decreases with increasing film thickness. For 4 ML, the lifetime of the (n=2) IPS electrons is even shorter than for electrons excited into the according (n=1) IPS. This differs strongly from the observations on insulating rare-gas layers with negative electron affinity (EA) on metals \cite{Berthold2004}. For such systems, the dielectric layer acts as a spacer and decouples the IPS from the metal surface. Due to the negative EA, no additional electronic states of the spacer layer are introduced in the energy range of the IPS. Hence, the wavefunctions of the IPS are pushed away from the metal surface. This results in a strong increase of the lifetimes with increasing layer thickness. The behavior of NiO films is, in contrast, similar to rare-gas adlayers with EA$\,>0$. For such systems \cite{Berthold2004}, the energetic position of the IPS is within the conduction band of the rare gas adlayers. Therefore, the wavefunction of the IPS can penetrate the layer. In case of bulk NiO, the conduction band minimum is located at 2.6 eV above $E_F$ \cite{Reinert1995}. The IPS of 2--4 ML NiO on Ag(001) have intermediate state energies of 3.7 eV and more and are, therefore, located in the NiO conduction band. Hence, the wavefunctions of the IPS can couple to those of the NiO thin film states. The decrease of lifetimes for the second IPS from 2 ML to 4 ML NiO/Ag(001) is explained by a stronger coupling of the IPS to thin film states due to an increase of the density of NiO states with increasing film thickness in the energy region of the IPS. 
    
    For all prepared oxide films, the feature B occurs at an intermediate energy of 3.9 eV. This state belongs to an electronic transition caused by the silver substrate and is not discussed further here.    
\section{Conclusions}
  We have investigated the image-potential states for NiO(001) ultrathin films of 2--4 ML on Ag(001) by means of angle- and time-resolved 2PPE. The first three IPS were identified via their parabolic dispersion and their relatively long lifetimes. The decay of the IPS depends strongly on the film quality. Compared to a bilayer NiO/Ag(001) prepared at RT, an annealed 2 ML film yield a factor of 1.3 and 1.7 longer lifetimes of the first and second IPS, respectively. For the first IPS the lifetimes are in the range of 27--42 fs and differ only slightly for annealed NiO films with 2--4 ML thickness. In contrast, the lifetimes of the second IPS decrease with increasing film thickness from 85 fs (2 ML) to 33 fs (4 ML). This is explained by the introduction of additional NiO thin film states in the energy range of the IPS. Therefore, electrons in the IPS decay faster due to an increased coupling to these states. 
\acknowledgments The authors thank Ralf Kulla for generous technical support. Financial support by the SFB 762 is gratefully acknowledged. K.G. thanks the International Max Planck Research School for Science and Technology of Nanostructures for funding.
\end{document}